\documentstyle[preprint,aps,prl]{revtex}
\begin{document}
\title{Continuum limit, Galilean invariance, and solitons in the
quantum equivalent of the noisy Burgers equation}
\author{Hans C. Fogedby $^1$}
\address{Institute of Physics and Astronomy,
University of Aarhus, DK-8000, Aarhus C, Denmark}
\author{Anders B. Eriksson $^2$ and Lev V. Mikheev$^3$}
\address{Nordita, Blegdamsvej 17, DK-2100, Copenhagen \O, Denmark}
\date{\today}
\maketitle
\begin{abstract}
A continuum limit of the non-Hermitian spin-1/2 chain, conjectured recently
to belong to the universality class of the noisy Burgers or, equivalently,
Kardar-Parisi-Zhang equation, is obtained and analyzed. The Galilean
invariance of the Burgers equation is explicitly realized in the operator
algebra. In the
quasi-classical limit we find nonlinear soliton excitations exhibiting the
$\omega\propto k^z$ dispersion relation with dynamical exponent $z=3/2$.
\end{abstract}
\draft
\pacs{
E-mail:\\
$1$. fogedby@dfi.aau.dk\\
$2$. eriksn@nordita.dk\\
$3$. mikheev@nordita.dk}
The Kardar-Parisi-Zhang (KPZ) equation plays an important role in modern
non equilibrium
statistical mechanics as a continuum description of growing interfaces
\cite{family,kpz}. In a co-moving frame the equation of motion
for the relative height variable $h(x,t)$ has the form
\begin{equation}
\partial h/\partial t=\nu\nabla^2h+\frac{1}{2}\lambda(\nabla h)^2+\eta \ ,
\label{kpz}
\end{equation}
where $\nu$ is an effective diffusion coefficient, $\lambda$  a
coupling constant characterizing the slope dependence of the
growth velocity, and $\eta(x,t)$ a white noise,\linebreak
$<\eta(x,t)\eta(x',t')>=\Delta\delta(x-x')\delta(t-t')$, representing
fluctuations either in the drive or in the environment.

The slope variable $u=\nabla h$ satisfies the noisy
Burgers equation
\cite{burgers,Forster et al}
\begin{equation}
\partial u/\partial t-\lambda u\nabla u=\nu\nabla^2u+\nabla\eta  \ .
\label{burgers}
\end{equation}
The universality class represented by (\ref{kpz}) and (\ref{burgers}),
besides driven interface and irrotational fluid dynamics,
includes such important models
as a driven lattice gas \cite{van Beijeren},
as well as a directed polymer \cite{Huse Henley},
or a quantum particle \cite{KardarReview}, in random environment.
These models serve as paradigms in the
theories of driven and disordered systems. An
additional attraction of the problem is its analytical tractability.
A consistent, although uncontrolled, perturbative
renormalization group flow has been constructed up to the two-loop level
\cite{Forster et al,Medina et al,Frey et al} and appears
to give qualitatively correct results. What emerges is a critical
(massless) behavior, which at and below two spatial dimensions is always
governed by a strong-coupling renormalization
group fixed point. Two basic scaling dimensions describe the
long-wave-length low-frequency behavior of the problem: i) the roughness
exponent $\zeta$ characterizing the height correlations in (\ref{kpz}) and
ii) the dynamical exponent $z$. The velocity field $u$ or, equivalently,
slope field $\nabla h$ for the interface, has scaling dimension
$1-\zeta$. The pair correlation function scales as
\begin{equation}
<u(x_0,t_0)u(x_0+x,t_0+t)>=x^{-2(1-\zeta)}f(t/x^z) \ ,
\end{equation}
where the scaling function $f(x)\propto x^{-2(1-\zeta)/z}$ for
large values of its argument.

The case  one spatial and one time dimension, $d=1+1$, is of particular
interest. Although
the perturbative renormalization group flows into the strong coupling
regime, outside of its range of validity,
it was noticed in \cite{Forster et al} that both scaling
dimensions follow explicitly from
two exact scaling relations. The first one follows from the
Galilean invariance of the original Burgers equation
(generalized to $\lambda\neq -1$), i.e.,
$u(x,t)\rightarrow u(x-\lambda u_0t,t)-u_0$,
leaving all averages unchanged. This property implies that
$\lambda$ is a structural constant
of the symmetry group and therefore remains invariant under a
renormalization group
transformation. Comparing the scaling properties of the two terms
on the left hand side of
(\ref{burgers}) we then obtain the first scaling relation, $z=2 -\zeta$.
The other scaling relation
is particular to the case $d=1+1$ and follows from the fact that the
simple Gaussian {\em equal time} stationary probability distribution,
$\ln P(u)\propto -\int u^2(x)dx$, satisfies
the Fokker-Planck equation for the probability distribution $P(u,t),$
\begin{eqnarray}
\frac{\partial P(u,t)}{\partial t} &&=
 -\int dx \frac{\delta}{\delta u}
\left [(-\lambda u
\frac{\partial u}{\partial x}
 + \nu \frac{\partial^2 u}{\partial x^2})
P(u,t)\right ] \nonumber \\
&& + \frac{\Delta}{2}\int dx dx'
\left [\frac{\partial^2}{\partial x \partial x'}\delta (x-x')\right ]
\frac{\delta^2 P(u,t)}{\delta u
\delta u'} \ ,
\label{fokker}
\end{eqnarray}
corresponding to (\ref{burgers}) for {\em any} value of $\lambda$.
This property then implies $\zeta=1/2$ (random walk behavior)
and thus $z=3/2$.

The exact rational values of the scaling exponents found in
$d=1+1$ bear a superficial resemblance
to the exponents encountered
in two-dimensional classical critical
problems, equivalent to
$d=1+1$ relativistic quantum field theories \cite{cardy}. Note, however,
that the value
$z=3/2$ is manifestly non-relativistic, making extensions of the
conformal field
theory methods a highly nontrivial task.
We might, however, still expect to gain new insight by studying the
equivalent
(1+1)-dimensional {\em nonrelativistic} quantum system by the
methods developed
for one-dimensional
spin chains \cite{mattis}.
Here spin waves and solitons in, e. g.,  the Heisenberg ferromagnet,
are characterized by $\omega=k^z$ with $z=2$
\cite{fogedby}, while activational quantum dynamics of the Ising chain in
a random transverse field implies $z=\infty$ \cite{dsf}.

Such a quantum spin chain approach has been proposed in
\cite{Gwa Spohn}, and
considered further in \cite{deNijs}, on the basis of the equivalence
between the master
equation describing the evolution of a one-dimensional
driven lattice gas or the
equivalent lattice interface growth model and the non-Hermitian
spin $s=1/2$  Hamiltonian for $N$ spins,
\begin{equation}
\hat{H} = -\sum_{j=1}^N \left[ \vec{S}_j\cdot\vec{S}_{j+1} - \frac{1}{4} +
i\vec{\epsilon}\cdot( \vec{S}_j\times\vec{S}_{j+1} )\right] \ .
\label{chain}
\end{equation}
The mapping is achieved by identifying the eigenvalues of the
$z$-projections of the spins, $S^z_j$, with the slope variable $u_j$
of the discrete interface model or the occupation numbers in the
lattice gas representation.
The interface dynamics is governed by two basic rates,
$r_\uparrow$
and $r_\downarrow$, corresponding to flipping ``up'' or ``down''
a kink
(a pair of neighboring interface segments with opposite slopes);
these flips map onto the spin exchange processes in (\ref{chain}).
The vector $\vec{\epsilon}$ is oriented along the $z$-axis and absorbing
the mean rate $r_\uparrow +r_\downarrow$ by rescaling time we have
$|\vec{\epsilon}|=
\epsilon\equiv(r_\uparrow-r_\downarrow)/(r_\uparrow +r_\downarrow)$,
measuring the strength of the drive.
For $\epsilon =0$
``up'' and ``down'' flips
are equally probable, and we obtain
the Heisenberg ferromagnet, with the dynamical exponent $z=2$,
corresponding to the noisy linear diffusion equation or the
Edwards-Wilkinson equation for interface dynamics
\cite{family,deNijs}. The
asymmetric limit of only ``up'' or only ``down'' flips,
$\epsilon =\pm 1$, has been solved exactly by the Bethe-ansatz method
\cite{Gwa Spohn},
and the finite-size gap in the spectrum has been shown to scale with the
length of the chain to the power $3/2$, in agreement with $z=3/2$
for the noisy Burgers universality class.

In this Letter we obtain the continuum  limit of the spin chain
(\ref{chain}) corresponding to the noisy Burgers universality class. We
show that
properly defined this limit exhibits Galilean symmetry as realized by
the algebra of generators of the continuous global transformations.
It is known that
in the Heisenberg ferromagnet \cite{fogedby}, as well as in other
$XYZ$ spin-1/2 chains \cite{luther}, quasi-classical
quantization of the classical solutions of the continuum equations of
motion correctly reproduces the low-energy sector of the Bethe-ansatz
solution.
Motivated by this fact we also  study the quasi-classical limit.
We find that the classical equations of motion have
exact solitary wave solutions, which after quantization form a
branch of elementary excitations with the $\omega\propto k^{3/2}$
dispersion relation.

As a guide to the proper continuum limit we make use of the two exact
relationships
employed above in the derivation of the scaling dimensions $\zeta$ and $z$.
The implementation of the second one is readily carried out by noting that
the stationary equal time probability distribution of the Burgers equation
maps onto the ground state of the quantum problem. The ground state of the
Heisenberg ferromagnet, $|0\rangle$, for $\epsilon=0$ is aligned
with maximum total spin $sN$
and fully degenerate spin direction.
In the following we quantize along the $x$-direction, i.e.,
$\langle 0|S^x_j|0\rangle=s$, $\langle 0|S^z_j|0\rangle=0$, and
$\langle 0|S^y_j|0\rangle=0$,
corresponding to a horizontal interface with zero slope.
With this choice
(cf. \cite{Gwa Spohn}) $S^z_j$ is
completely disordered and the correlation function
$\langle 0|S^z_{j}S^z_{j'}|0\rangle=s^2\delta_{j,j'}$.
In the continuum limit $\delta_{j,j'}/a\rightarrow\delta(x)$, where $a$
is the lattice spacing, implying the rescaling
$S^z(x)\sim S^z_j/a^{1-\zeta}$ with $\zeta=1/2$.
This result is also obtained by noting that a block
spin $S^z(x)$ constructed from a random sequence of $S_i=\pm 1/2$
yields a spin amplitude
of the order of the square root of the block size. Finally, in
the mapping of a master equation onto an equivalent quantum Hamiltonian
the energy levels correspond to the relaxation rates.
The ground state maps onto the stationary distribution,
consequently, its energy must
be equal to zero. It is easily seen that the aligned
ground state $|0\rangle$ is also an eigenstate of $\hat{H}$ in
(\ref{chain}) with eigenvalue zero, owing to the
cross product form of the last term. Consequently,
the ground state structure and thus the scaling
dimension of the spin density is not changed in the presence of the drive.

The excitation spectrum, however, is altered since we expect the
$\omega\propto k^2$ dispersion law of the Heisenberg ferromagnet to change
to $\omega\propto k^{3/2}$ in the presence of the drive term.
Clearly, the Galilean invariance which permits the evaluation
of the exponent $z$
requires the continuum limit $a\rightarrow 0$
to be taken since
discrete distances cannot be mixed with continuous time.

With the ground state aligned in the $x$-direction the excitations
above the ground state correspond to spin fluctuations in the $y$ and
$z$-directions. Noting that a block spin pointing in the $x$-direction
has a size $s\sim a^{-1}$ for $a\rightarrow 0$ while the spins in
the $y$ and $z$ directions scale as $a^{-1/2}$, we infer that the
continuum limit is equivalent to the large spin limit in the
block spin representation. Using the quasi-classical harmonic
oscillator representation
for the spin operators, see e.g. \cite{mattis}, we have
\begin{eqnarray}
S_j^x = s-\frac{1}{2}(\hat{u}_j^2+\hat{\varphi}_j^2)\ ,\nonumber  \\
S_j^y = \hat{\varphi}_j s^{1/2},\  S_j^z = \hat{u}_j s^{1/2} \ .
\end{eqnarray}
The oscillator position and momentum operators $\hat{\varphi_j}$
and $\hat{u}_j$, satisfy the commutator relation
$[\hat{\varphi_j},\hat{u}_k]=i\delta_{jk}$ and correspond to the
polar angle and the $z$-component of the spin, i.e., the action-angle
representation of the Heisenberg spin \cite{fogedby}.
In the
continuum limit the field operators
$\hat{u}(x)$ and $\hat{\varphi}(x)$ absorb the scale factor $a^{1/2}$,
$\hat{u}(x)=\hat{u}_ja^{-1/2}$ and
$\hat{\varphi}(x)= \hat{\varphi}_ja^{-1/2}$,
and obey the canonical commutation relation
$[\hat{\varphi}(x),\hat{u}(x')] = i \delta(x-x')$. Introducing the
spin-wave
stiffness $J=a^{1/2}$, using periodic boundary conditions, and
omitting constant
terms, the Hamiltonian finally reads
\begin{equation}
\hat{H} = a^{3/2}\int dx \left[ \frac{J}{2}\left( (\frac{\partial
\hat{u} }{\partial x})^2
+(\frac{\partial \hat{\varphi} }{\partial x})^2\right) +
i\epsilon \hat{u}^2 \frac{\partial \hat{\varphi}}{\partial x} \right] \ ;
\label{continuum}
\end{equation}
we note that the overall factor $a^{3/2}$ is consistent with
the dynamical exponent $z=3/2$.
It is easily seen that the first term in (\ref{continuum})
corresponds to the spin-wave
approximation for the Heisenberg ferromagnet
(the first term in (\ref{chain})); the second term representing the
drive then corresponds to a
spin-wave interaction.

We can now show that the Hamiltonian (\ref{continuum}) is
equivalent to the Fokker-Planck equation (\ref{fokker})
for the noisy Burgers equation (\ref{burgers}).
Introducing the conjugate
operators $u$ and $i\delta/\delta u$ and performing the canonical
transformation
$i\delta/\delta u = (\nu/\Delta)^{1/2} \left [
\hat{\varphi}(x) - i \hat{u}(x) \right ]$ and
$u = (\Delta/\nu)^{1/2} \hat{u}(x)$
the operator ${\cal L}$ in
$dP(u,t)/dt ={\cal L}(u) P(u,t)$, Eq.\ (\ref{fokker}) is equivalent to
(\ref{continuum}) for
$J=\nu$ and $\epsilon=\frac{1}{2}\lambda(\Delta/\nu)^{1/2}$.

Noting that the time evolution operator for the master
equation
is $\exp(-t\hat{H})$ \cite{Gwa Spohn}, the equations of motion
for $\hat{\varphi}$ and $\hat{u}$
follow from $d\hat{\varphi}/dt=[\hat{H},\hat{\varphi}]$ and
$d\hat{u}/dt=[\hat{H},\hat{u}]$,
\begin{eqnarray}
\frac{\partial \hat{\varphi}}{\partial t} &=& i J
\frac{\partial^2 \hat{u}}{\partial x^2} + 2 \epsilon \hat{u}
\frac{\partial \hat{\varphi}}{\partial x} \ , \\
\frac{\partial \hat{u}}{\partial t} &=& -i J
\frac{\partial^2 \hat{\varphi}}{\partial x^2} + 2 \epsilon \hat{u}
\frac{\partial \hat{u}}{\partial x} \  .
\end{eqnarray}
These field equations are invariant under the
Galilean transformation
\begin{equation}
\left\{ \begin{array}{l}
\hat{u}(x,t) \to \hat{u}(x-2\epsilon u_0t,t) - u_0 \ , \\
\hat{\varphi}(x,t) \to \hat{\varphi}(x-2\epsilon u_0t,t) \ ,
\label{galilean2}
\end{array}\right.
\end{equation}
in addition to being invariant under an arbitrary shift in $\hat{\varphi}$.
These two continuous
symmetry relations correspond to the two generators of the
full group of rotations of the Heisenberg spin in (\ref{chain})
for $\epsilon=0$. The oscillator representation
corresponds to a local Abelian limit of the
non-Abelian group $O(3)$; the
spin wave Hamiltonian is invariant under arbitrary shifts in
both canonical fields $\hat{\varphi}$ and
$\hat{u}$. In the presence of the spin wave interaction,
the invariance under
shifts in $\hat{\varphi}$ is preserved
while the other symmetry is replaced by the Galilean
space-time mixing
(\ref{galilean2}).

These symmetries have a compact representation in terms of the
algebra of the
operators $\hat{M}^z=\int dx \hat{u}$ and
$\hat{\Phi}=\int dx \hat{\varphi}$ generating rotations
about the $y$ and
the $z$-axis, or, equivalently, shifts in $\hat{\varphi}$
and $\hat{u}$, respectively. Adding the momentum operator
$\hat{P}=\int dx \hat{u} \partial/\partial\hat{\varphi}$ and $\hat{H}$,
generating  space and time translations,
respectively, we arrive at the operator algebra:
$[\hat{H},\hat{M}^z] = [\hat{H},\hat{P}] = [\hat{P},\hat{\Phi}] =
[\hat{P},\hat{M}^z] = 0$,
and $[\hat{\Phi},\hat{M}^z] = i L$ , where $L$ is the length of
the interface, while the Galilean invariance yields
\begin{equation}
\label{eq:commut}
[\hat{H},\hat{\Phi}]= -2i \epsilon \hat{P} \ .
\end{equation}
Thus $\epsilon$ is
indeed a structural constant of the algebra of symmetries.
This justifies the renormalization leading to (\ref{continuum})
which were chosen so that $\epsilon$ is invariant under changes in the
microscopic scale $a$.
The above commutation relation has a simple
interpretation in the case of the elementary excitations, the lowest
energy states carrying momentum $k$; (\ref{eq:commut})
 implies rotation of the
global polar angle $\hat{\Phi}$ with constant angular
velocity $d\hat{\Phi}/dt=2\epsilon k$.

We now attempt to find an elementary excitation with the
$\omega\propto k^{3/2}$ dispersion relation by a direct analysis of
the nonlinear equations of motion (8) and (9).
Such an analysis is most easily carried out in the classical limit
followed by quasi-classical quantization (cf. \cite{fogedby}).
We therefore first replace the operators $\hat{\varphi}$
and $\hat{u}$ in (\ref{continuum}) by the
field variables $\varphi$ and $u$ satisfying
the Poisson bracket $\{ u(x),\varphi(y)\} =\delta(x-y)$. The classical
equations of
motion $\partial u/\partial t=-i\{ H,u\}$ and
$\partial \varphi/\partial t=-i\{ H,\varphi\}$ then have the same form
as (8) and (9) with $\hat{u}$ and
$\hat{\varphi}$ replaced by $u$ and $\varphi$.
For $\epsilon=0$ we find small amplitude spin waves about the
ground state with quadratic dispersion leading to $z=2$. However, for
$\epsilon\neq 0$ we identify non-linear non-perturbative soliton
solutions propagating with
velocity $v$, i.e., replacing
$\partial/\partial t$ by $-v\partial/\partial x$
we obtain $-v\partial u/\partial x=-iJ\partial^2\varphi/\partial x^2
+2\epsilon u \partial u/\partial x$ and
$-v\partial\varphi/\partial x=
iJ\partial^2 u/\partial x^2
+2\epsilon u\partial\varphi/\partial x$
which are easily solved by quadrature.

However,
for our purposes we note that by forming the quotient of
the two equations
and integrating once we obtain
$(\partial u/\partial x)^2+(\partial\varphi/\partial x)^2=C$.
For solutions with vanishing derivatives
at infinity $C=0$,
and we have
\begin{equation}
\label{eq:u_phi_rel}
\partial u/\partial x=\pm i\partial\varphi/\partial x \ .
\end{equation}

By insertion we obtain one equation for the slope $u$
which by quadrature yields
$-vu=\pm J\partial u/\partial x+\epsilon u^2+ B$
Relating $B$ to the
boundary values $u_{\pm}=u(\pm\infty)$ far away from the soliton
we obtain $\partial u/\partial x
=\pm(\epsilon/J)(u-u_+)(u-u_-)$
and the ``mean amplitude--velocity'' condition
\begin{eqnarray}
u_+ + u_-=-v/\epsilon\ ,
\label{bound}
\end{eqnarray}
defining permanent profile kink or soliton solutions.
The soliton shape is given by
$|(u-u_+)/(u-u_-)|=
\exp{[\pm(\epsilon/J)(u_+-u_-)(x-x_0)]}$
with ``center of mass'' position $x_0$. For
$\partial u/\partial x=\pm i\partial\varphi/\partial x$
we have solitons (kinks) with positive and negative slopes, respectively.
In order to interpret these solutions as excitations above the ground
state, which classically is the uniform solution $u=0$, we have to satisfy
the boundary conditions $u(\pm\infty)=0$. We can achieve this by grouping
kinks in pairs: for $v>0$ we group the $(u_-,u_+)=(0,-v/\epsilon)$ kink at
smaller $x$ with
$(u_-,u_+)=(-v/\epsilon,0)$ at larger $x$;  for $v<0$ the
$(u_-,u_+)=(0,v/\epsilon)$ kink is followed by $(u_-,u_+)=(v/\epsilon,0)$.
The distance between
the kinks in a pair is much larger than the intrinsic width $J/v$
but much smaller than the length of the system.

The energy momentum relationship for a pair is obtained by noting
that the energy $E$ is given by (\ref{continuum}) while the momentum,
the generator of translation, is given by
$P=-\int dx u \partial\varphi/\partial x$.
Inserting (\ref{eq:u_phi_rel}) we
obtain for a single kink, $E=\pm (1/3)a^{3/2}(u_+^3-u_-^3)$ and
$P=\pm (1/2)(u_+^2-u_-^2)$. Note that the two kinks in each pair
moving with a constant velocity $v$
correspond to the opposite choices of sign in
(\ref{eq:u_phi_rel});
as a result both energy and momentum are doubled and we obtain
\begin{equation}
E=\frac{2}{3}a^{\frac{3}{2}}\left (\frac{v}{\epsilon}\right )^3\ , \
P=\pm i\left (\frac{v}{\epsilon}\right )^2
\label{ener-mom}
\end{equation}
for $v> (<)\ 0$.
We note that the momentum is purely imaginary; this is a feature of
the complex field equations (8) and (9). Eliminating
the amplitude $v/\epsilon$ we obtain the dispersion law
\begin{eqnarray}
E\propto a^{\frac{3}{2}}\left (\pm iP\right )^{\frac{3}{2}}.
\label{disp-rel}
\end{eqnarray}
In order to revert to a discussion of the quasi-classical limit of the
spin chain the final step in our analysis amounts to a quasi-classical
quantization of the above pair solutions. In the spirit of the Landau
quasi-particle picture we envisage that we can label the low lying
quantum states in terms of a dilute gas of pairs of non-linear
kinks with dispersion law (\ref{disp-rel}).

A few comments are in order:
a) The amplitude of a pair, $v/\epsilon$,
diverges for $\epsilon\rightarrow 0$ implying that these are indeed
non-perturbative solutions of the field equations characterizing the
strong coupling renormalization group fixed point. This is also consistent
with the fact that the coupling $\epsilon$ does {\em not} appear
in the dispersion law (\ref{disp-rel}), displaying the universality of the
strong coupling regime.
b) The kink solutions characterized by (\ref{bound}) are manifestly
invariant under the Galilei transformation (\ref{galilean2}).
c) Inserting $\partial \varphi/\partial x=i\partial u/\partial x$
in the field equations (8) and (9) we obtain the deterministic
noiseless Burgers equation. We conclude that the above kink solutions
correspond to the well-known shock wave solutions \cite{burgers}.
In the interface representation a pair of kinks  forms a {\em step}:
a localized slope fluctuation propagating in
the lateral direction.

In conclusion, two main results for the theory
of the noisy Burgers-KPZ universality class
have been obtained in this Letter.
First, although supported by an exact
Bethe ansatz solution for a special value of the parameter
$\epsilon$ and by numerical results, the
equivalence between the lattice model (\ref{chain})
and the continuum Langevin equations (1) and (2)
remained a conjecture \cite{Gwa Spohn,deNijs}.
Here we have shown, using well-known methods in the quantum theory
of spin chains \cite{mattis}, how to implement
the continuum limit of the lattice model (\ref{chain}). The rescaling
 leads to
a continuum Hamiltonian (\ref{continuum})
which by a canonical transformation
is reduced to the Liouville operator (\ref{fokker})
of the Fokker-Planck equation corresponding to
the noisy Burgers-KPZ equations. The field operators
appearing in the continuum Hamiltonian absorb the
powers of the lattice constants making their
dimensions equal to the scaling dimensions of
the corresponding quantities of the noisy
Burgers-KPZ universality class.
The choice of the scaling
dimensions is governed by the (generalized)
Galilean invariance of the theory (\ref{galilean2}), which is
explicitly realized in the commutation relation (\ref{eq:commut})
of the symmetry algebra.

Second, the space-time scaling $t\propto x^z$
with the exponent
$z=3/2$ appears in the theory of the $d=1+1$
noisy Burgers equation as the only one consistent with
the Galilean symmetry of the problem
\cite{Forster et al}. However, little physical
intuition exists concerning the physical realization of
this dynamical scaling. Here we
have made a step towards a better understanding
of the physical nature of the problem by identifying
explicit nonlinear soliton solutions to the
classical limit of the nonrelativistic field
theory. These solitons correspond to the
shock wave solutions of the {\em deterministic}
Burgers equation. Grouping shock waves of the opposite
sign moving with the same velocity, we obtain a localized
excitation with respect to a uniform solution,
which in the KPZ representation corresponds
to a step propagating along a growing interface.
In the quasi-classical limit these soliton pairs
form a branch of quantum excitations, exhibiting
the nontrivial $\omega\propto k^{3/2}$ dispersion
relation.

Discussions with Jari Kinaret, Alan Luther,
and Alexander Nersessyan are gratefully acknowledged.

\end{document}